\documentclass[oldversion,PRINTER,letterpaper]{aa}
\usepackage{natbib}
\usepackage{graphicx}
\begin{document}

\title{Properties of simulated sunspot umbral dots}

\author{L. Bharti \and B. Beeck \and M. Sch\"ussler}

\institute{Max-Planck-Institut f\"ur Sonnensystemforschung\\
       Max-Planck-Str.~2, 37191 Katlenburg-Lindau, Germany\\
\email{bharti@mps.mpg.de}}

\date{\today}

\abstract{ Realistic 3D radiative MHD simulations reveal the
magneto-convective processes underlying the formation of the
photospheric fine structure of sunspots, including penumbral filaments
and umbral dots. Here we provide results from a statistical analysis of
simulated umbral dots and compare them with reports from
high-resolution observations. A multi-level segmentation and tracking
algorithm has been used to isolate the bright structures in synthetic
bolometric and continuum brightness images. Areas, brightness, and
lifetimes of the resulting set of umbral dots are found to be
correlated: larger umbral dots tend to be brighter and live longer. The
magnetic field strength and velocity structure of umbral dots on
surfaces of constant optical depth in the continuum at 630~nm indicate
that the strong field reduction and high velocities in the upper
parts of the upflow plumes underlying umbral dots are largely hidden
from spectro-polarimetric observations. The properties of the simulated
umbral dots are generally consistent with the results of recent
high-resolution observations. However, the observed population of small,
short-lived umbral dots is not reproduced by the simulations, possibly
owing to insufficient spatial resolution.  \keywords{Sun: magnetic
fields - Sun: photosphere - sunspots} } \maketitle

\section{Introduction}

Recent realistic MHD simulations have revealed that the complex
brightness and flow structure of sunspots results from convective energy
transport dominated by a strong vertical (in the umbra) or inclined (in
the penumbra) magnetic field \citep{Heinemann:etal:2007,
Scharmer:etal:2008, Rempel:etal:2009a, Rempel:etal:2009b}. In the umbra,
such magneto-convection occurs in the form of narrow upflow plumes with
adjacent downflows \citep[][henceforth referred to as
SV06]{Schuessler:Voegler:2006}.  In computed intensity images, these
plumes appear as bright features that share various properties with
observed umbral dots.  Here we provide a systematic study of the umbral
simulation results with specific emphasis on the properties of the
bright structures and their comparison with recent results from
high-resolution observations.

\section{MHD simulation}

We used the MURaM code \citep{Voegler:2003, Voegler:etal:2005} with
nearly the same simulation setup as SV06. The dimensions of the
computational box are $5.76\,$Mm $\times$ $5.76\,$Mm in the horizontal
directions and $1.6\,$Mm in the vertical. Rosseland optical depth unity
is about $0.4\,$Mm below the upper boundary. The grid cells have a size
of $20\,$km in the horizontal directions and $10\,$km in the vertical.
The magnetic diffusivity is 2.8$\cdot10^6\,$m$^2$s$^{-1}$, the lowest
value compatible with the grid resolution. The hyperdiffusivities for
the other diffusive processes \citep[for details,
see][]{Voegler:etal:2005} in the deepest layers of the box were chosen
lower than the values used by SV06 in order to minimize their
contribution to the energy transport. Since we do not synthesize
spectral lines or Stokes profiles, the radiative transfer can be treated
in the gray approximation. The fixed vertical magnetic flux through the
computational box corresponds to a mean field strength of
$2500\,$G. Side boundaries are periodic, and the top boundary is closed
for the flow.  The lower boundary is open and the thermal energy density
of the inflowing matter has been fixed to a value of
$3.5\cdot10^{12}\,$erg$\cdot$cm$^{-3}$, leading to a total surface
energy output comparable to that of a typical sunspot umbra, i.e., about
20\% of the value for undisturbed solar surface.  The magnetic field
is assumed to be vertical at the top and bottom boundaries.

We started our simulation from the last snapshot of the
run analyzed by SV06 and continued for 10.8 hours solar time in order to
obtain a sufficiently large statistical sample of simulated umbral dots.
Since the average magnetic field in the simulation is vertical, our
results represent the inner part of a sunspot umbra with central umbral
dots -- as opposed to peripheral umbral dots, which often are related to
penumbral filaments \citep[e.g.,][]{Grossmann-Doerth:etal:1986}.

\section{Image segmentation}

To obtain statistical properties of the large number of simulated umbral
dots (UDs), we used an automated detection method to distinguish between
bright features and the surrounding dark background (DB). Since both the
UDs and the background cover a broad range of intensities, we chose the
MLT (Multi Level Tracking) algorithm of \citet{Bovelet:Wiehr:2001} as
the basis of our image segmentation.  This method has been successfully
applied to various observational data \citep[e.g.,][]{Wiehr:etal:2004,
Bovelet:Wiehr:2007, Bovelet:Wiehr:2008, Kobel:etal:2009}, including
high-resolution images of sunspot umbrae
\citep{Riethmueller:etal:2008b}.

\begin{figure}
\centering
\resizebox{\hsize}{!}{\includegraphics{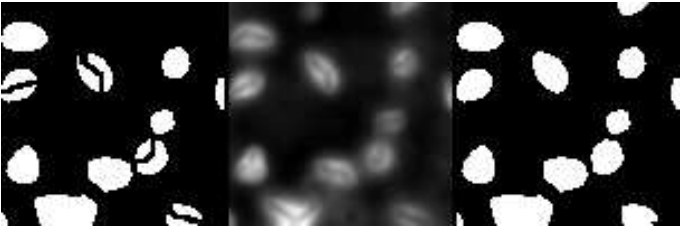}}
\caption{ Over-segmentation of simulated umbral dots due to
         dark lanes.  The section of the original image (middle panel)
         shows umbral dots with dark lanes. The unmodified MLT algorithm
         leads to an undesired splitting of many dots in the segmented
         image (left panel). Introducing a modification of the algorithm
         controlling the merging of segments removes the problem in
         most cases (right panel).}
\label{fig:DL_problem}
\end{figure}

The basic concept of the MLT algorithm is to segment the image by
applying a sequence of intensity thresholds with decreasing values,
keeping all features already detected.  In our case, a complication
arises from the dark lanes shown by many of our simulated UDs. This can
lead the algorithm to split an UD into two parts (see
Fig.~\ref{fig:DL_problem}).  In order to avoid such undesired splitting,
we permit the merging of segments that were separate for threshold $n$
after applying threshold level $n+1$ provided that their maximum
intensity does not exceed a value equal to the latter threshold
multiplied by a suitably chosen factor $c$. For our analysis, we
used 15 intensity thresholds ranging from about 2.5 to 1.5 times the
intensity of the dark background (somewhat depending on the dataset
used, see Sec.~4.2) and a value of $c=1.45$, which suppresses most cases
of unwanted splitting of UDs.

\begin{figure}
\centering
\resizebox{\hsize}{!}{\includegraphics{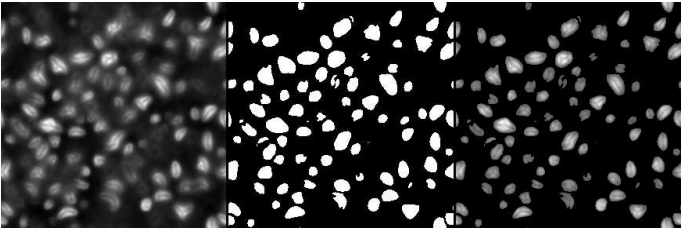}}
\caption{ Result of the MLT image segmentation. {\em Left:}
         original image; {\em Middle:} segmented image (binary map);
         {\em Right:} segmented umbral dot areas with their original
         intensities.}
\label{fig:segmentation}
\end{figure}

In a final step of the segmentation process, we removed from the
list of UDs all segmented structures whose maximum intensities do not
exceed the lowest threshold by at least 10\%; such features are
considered to be mere fluctuations of the background.
Fig.~\ref{fig:segmentation} shows an example of the resulting
segmentation of a typical simulated image.

\ifnum 2<1
In the best segmentation obtained so far, still remained some
structures, which do not look like UD's. These were very faint UD's not
wholly detected by the algorithm or no UD's at all. To get rid of these
things every region which maximum intensity does not exceed $4.45\cdot
10^9\,\frac{\text{erg}}{\text{cm}^2\text{s}\text{sr}}$ ($4.5\cdot
10^9\,\frac{\text{erg}}{\text{cm}^2\text{s}\text{sr}}$ in Data 1) is set
to zero! Fig. \ref{best} displays a sample of the original data (right)
segmented regions (middle), and the segmented regions refilled with the
original intensities. Some UD's still are effected by the DL-problem,
but applying much higher CP's would lead to merging UD's.\\
\fi

In order to study UD evolution and lifetimes, we extended the MLT
algorithm to include time, such that the segmentation can be carried
out in three dimensions (two spatial and one temporal) to follow the
temporal evolution of individual UDs.

\section{Results}

\subsection{Overall temporal evolution}
\label{subsec:temporal}

The relative inefficiency of the convective energy transport in the
strong umbral magnetic field leads to a rather long thermal relaxation
time of the system, so that it can take hours until it settles down
into a statistically stationary state. With our long simulation run of
10.8 hours solar time, we were able to follow the overall changes of the
system throughout the relaxation phase.

\begin{figure*}
\centering
\resizebox{\hsize}{!}
{\includegraphics{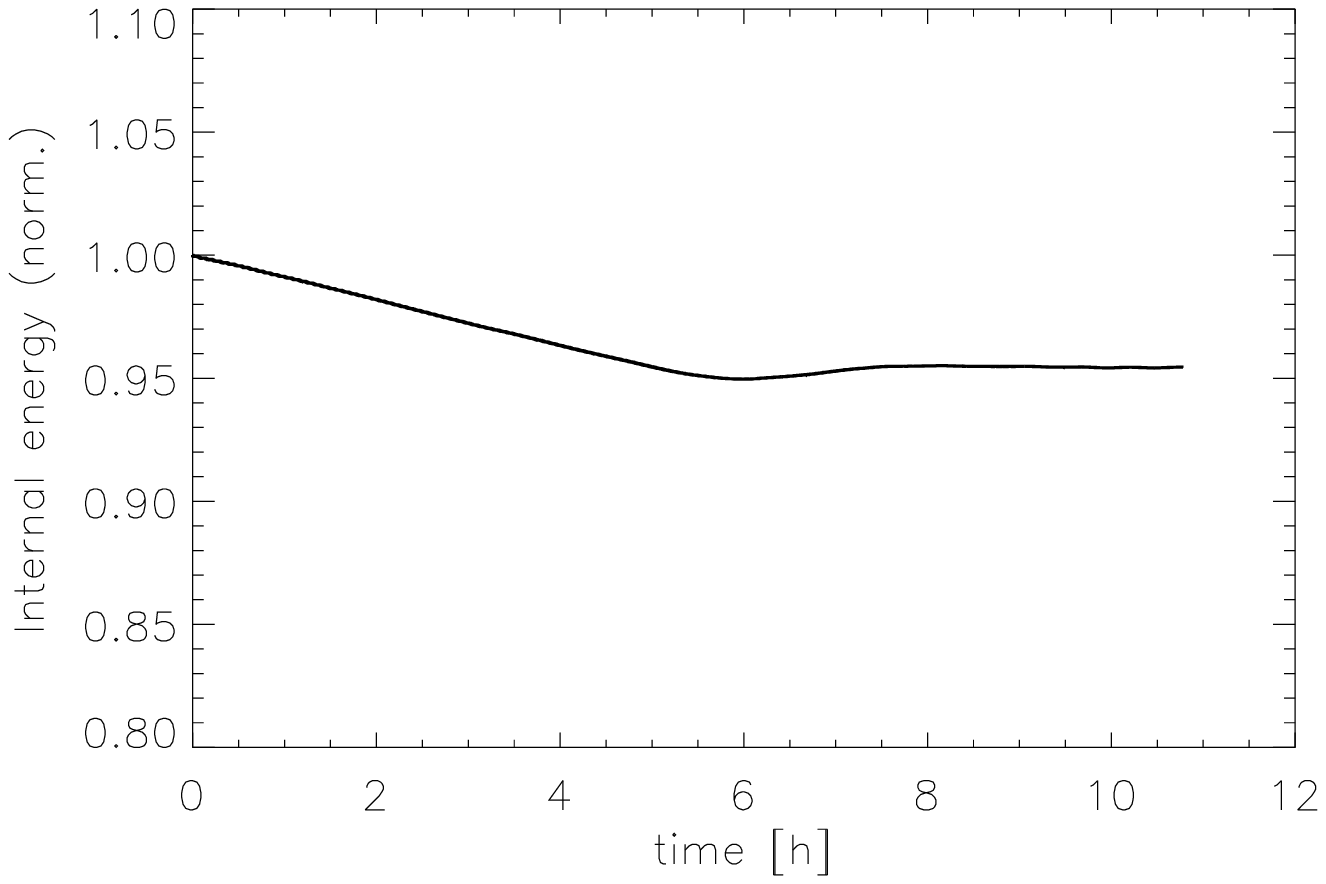}
\includegraphics{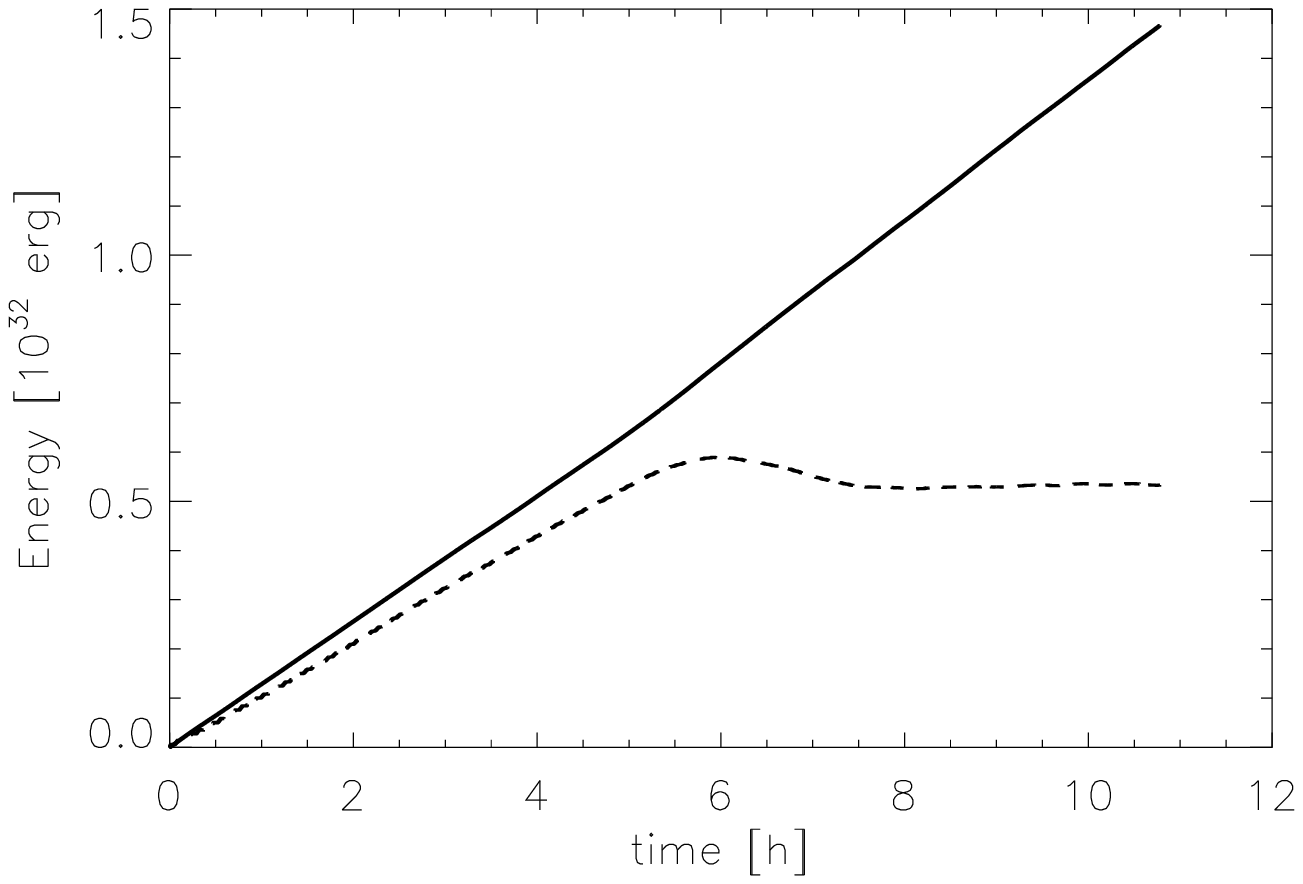}}
\caption{Thermal relaxation of the system. {\em Left
  panel:} time evolution of the total internal energy (normalized to its
  initial value); {\em Right panel:} total (radiative) energy output
  (solid line) and loss of total internal energy (dashed line), both
  integrated in time from $t=0$ onward.}
\label{fig:eint}
\end{figure*}

\begin{figure*}
\centering
\resizebox{\hsize}{!}
{\includegraphics{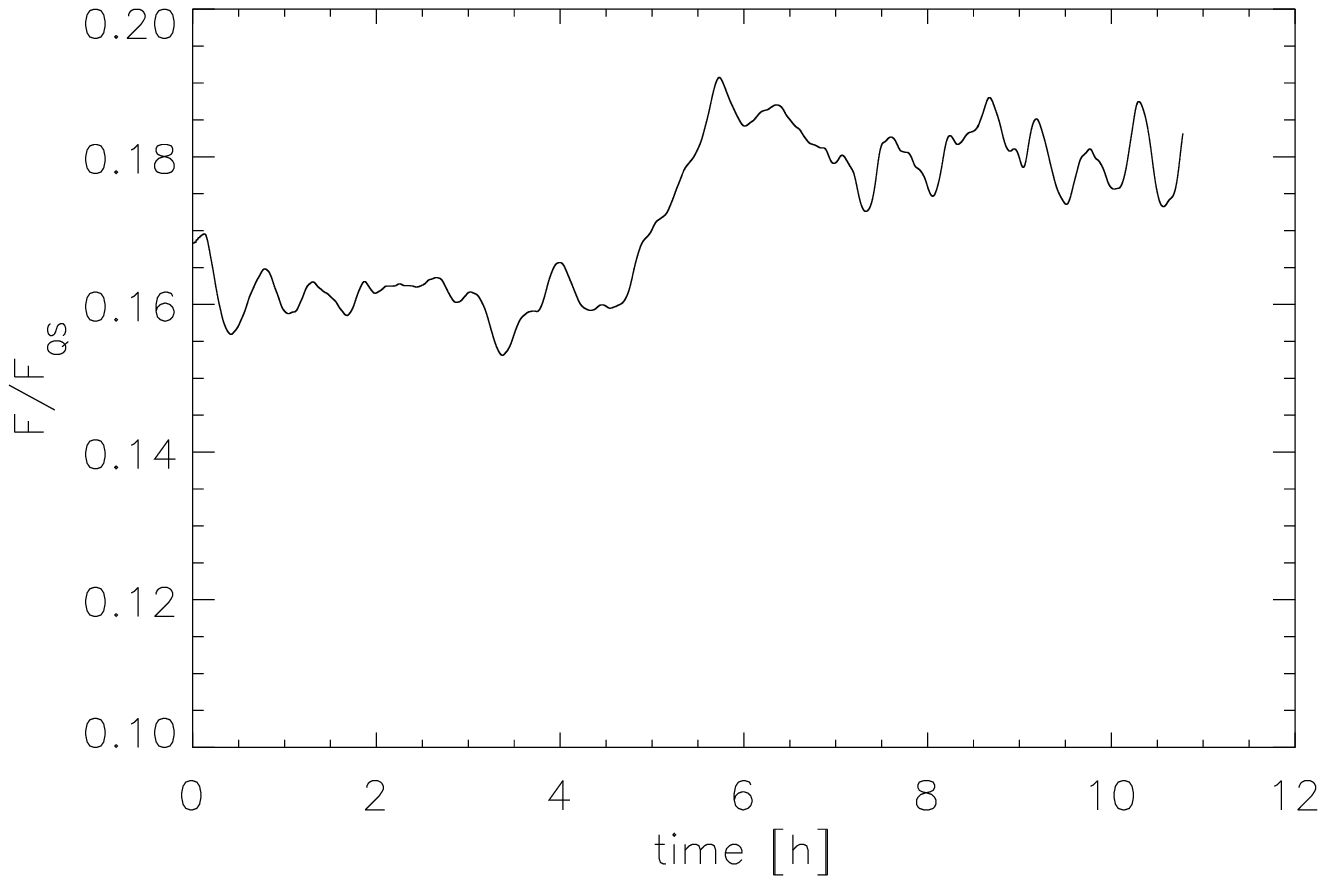}
\includegraphics{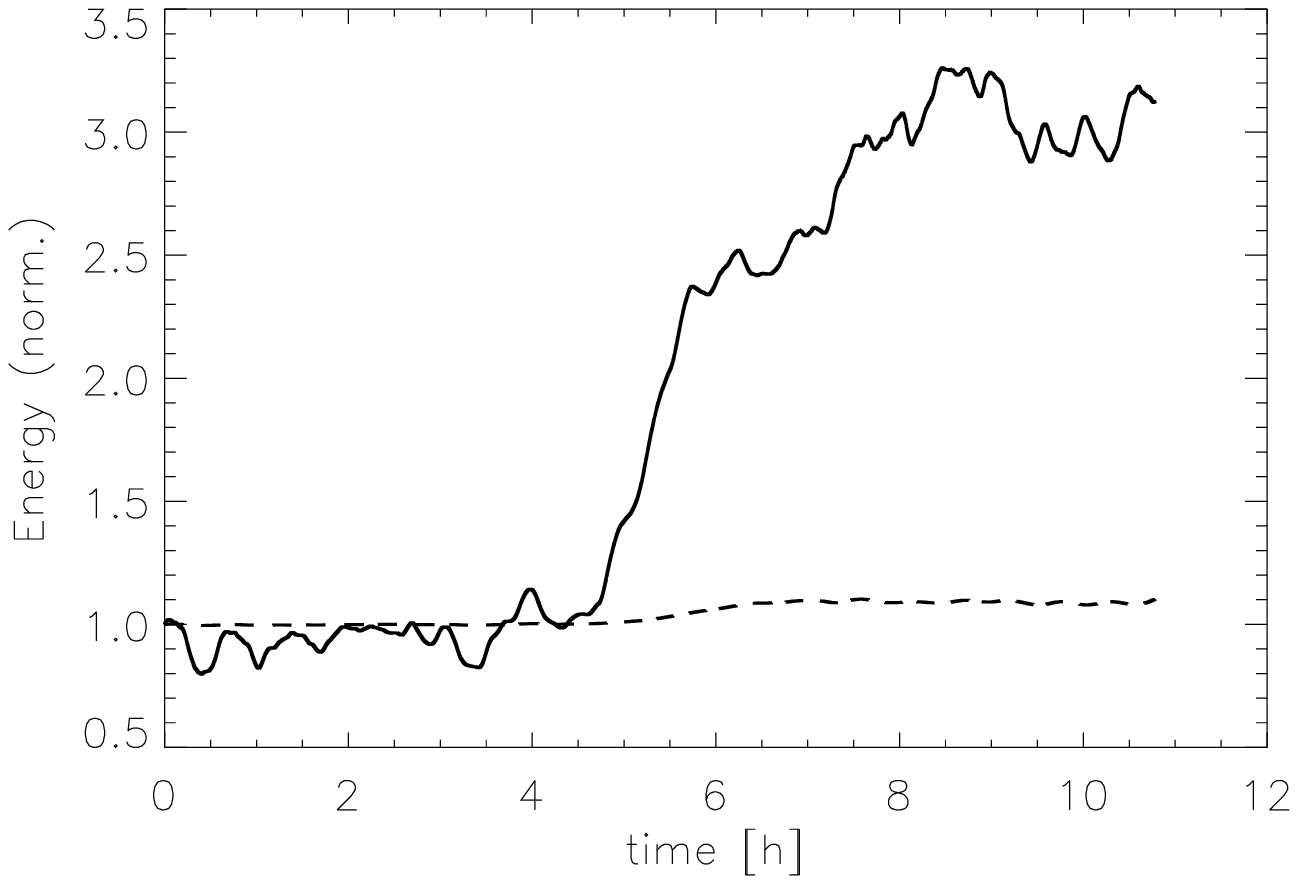}}
\caption{Temporal evolution of the system. {\em Left
  panel:} radiative energy flux from the box, normalized by the
  corresponding value for the same area of undisturbed solar
  photosphere; {\em Right panel:} total kinetic energy (solid line) and
  magnetic energy (dashed line), both normalized to their values at
  $t=0$.}
\label{fig:fout+ek+em}
\end{figure*}

The steady decrease of the total internal energy in the computational
box shown in Fig.~\ref{fig:eint} (left panel) clearly indicates that the
sytem is not relaxed until about 6--7 hours after the start of the
simulation.  Comparing the time-integrated total radiative output from
the box with the integrated decrease of the total internal energy given
in the right panel of Fig.~\ref{fig:eint} suggests that during the first
half of the evolution the energy output was mostly covered by a loss of
internal energy, i.e., the plasma in the box cooled. Only after about 6
hours, the two curves start to diverge and the internal energy does no
longer decrease.  At about the same time, the slope of the integrated
energy output steepens, indicating a somewhat higher energy output. This
is confirmed by Fig.~\ref{fig:fout+ek+em} (left panel), which shows that
the total radiative flux is higher by about 10\% in the second half of
the run.

These results suggest that the character of the energy supply has
changed halfway through the run: while the radiated energy was taken
from the internal energy in the first half and was not significantly
replenished through convective inflows through the bottom of the box,
convection became more effective in the second half of the run and led
to a thermally relaxed stationary situation with a somewhat higher total
radiative output. This interpretation is supported by considering the
time evolution of kinetic energy (Fig.~\ref{fig:fout+ek+em}, right
panel), which shows an increase by a factor of 3 in the second half of
the run. The magnetic energy also grows slightly, by about 10\%,
probably due to the stronger fluctuations (local compressions of the
magnetic flux) caused by the higher velocities.

\subsection{Statistical properties of umbral dots}
\label{subsec:propUD}

\begin{figure}
\centering
\resizebox{\hsize}{!}{\includegraphics{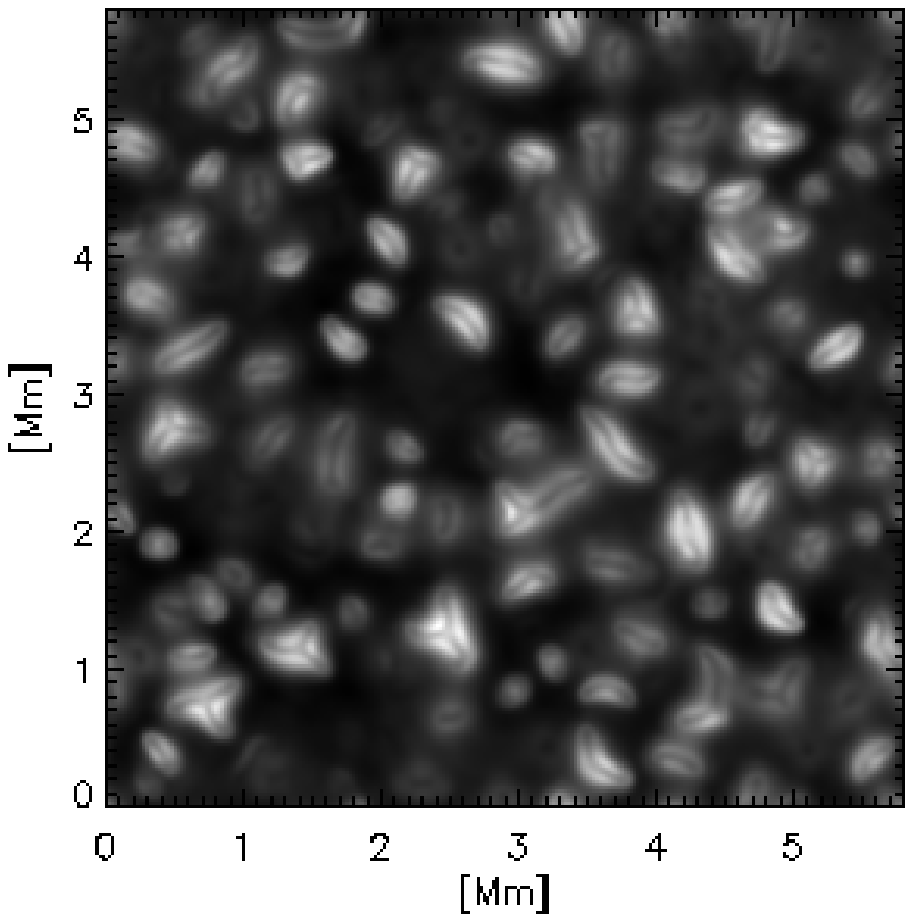}
\includegraphics{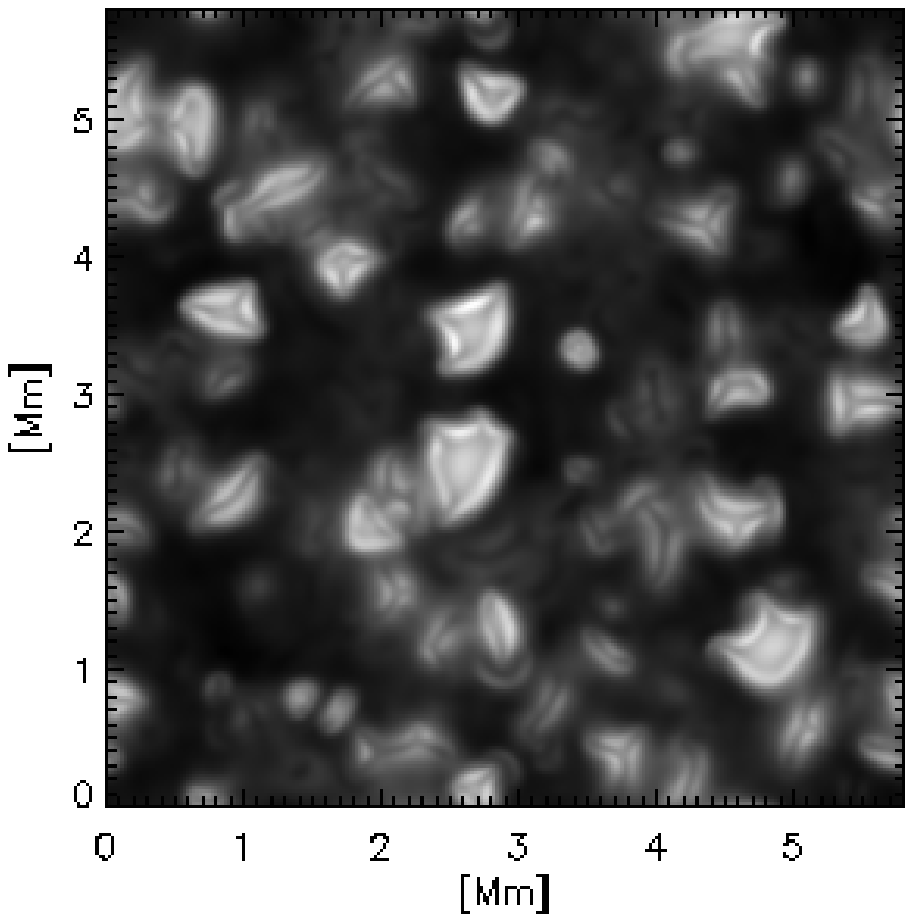}}
\caption{ Snapshots of the bolometric emergent intensity from
  the two phases of the simulation. {\em Upper panel:} initial cooling
  phase ($t\simeq 2\,$h); {\em Lower panel:} convective phase ($t\simeq
  9\,$h).}
\label{fig:snapshots}
\end{figure}

The existence of the two phases (cooling phase and convective phase) in
the temporal evolution of the umbra simulation is also reflected in
differences of the average UD properties in both phases. The snapshots
of (bolometric) brightness in Fig.~\ref{fig:snapshots} indicate that
larger UDs appear in the second phase while the total number of UDs
decreases.  We therefore defined separate datasets for each phase
and analyzed them individually. The first part (P1) covers the time
between 0.5~h and 4.5~h after the beginning of the run while the second
part (P2) extends from 6.2~h to 10.3~h. We excluded the transition
between the two phases and only kept periods for which the average UD
properties do not show significant secular trends.

For the MLT segmentation considered in this section, we used maps
of the bolometric (frequency-integrated) emergent intensity in the
vertical direction. The maps are separated in time by about 1 min each,
so that UDs can be analyzed in a statistical manner as well as their
evolution followed in time. Owing to the different characteristics of
the UDs, slightly different MLT threshold levels were chosen for P1
and P2, respectively. For dataset P1, the 15 intensity thresholds range
between 20\% and 31\% of the average quiet-Sun brightness while for P2
they range from 22\% to 35\%.

\begin{figure}
\centering
\resizebox{1.\hsize}{!}{\includegraphics{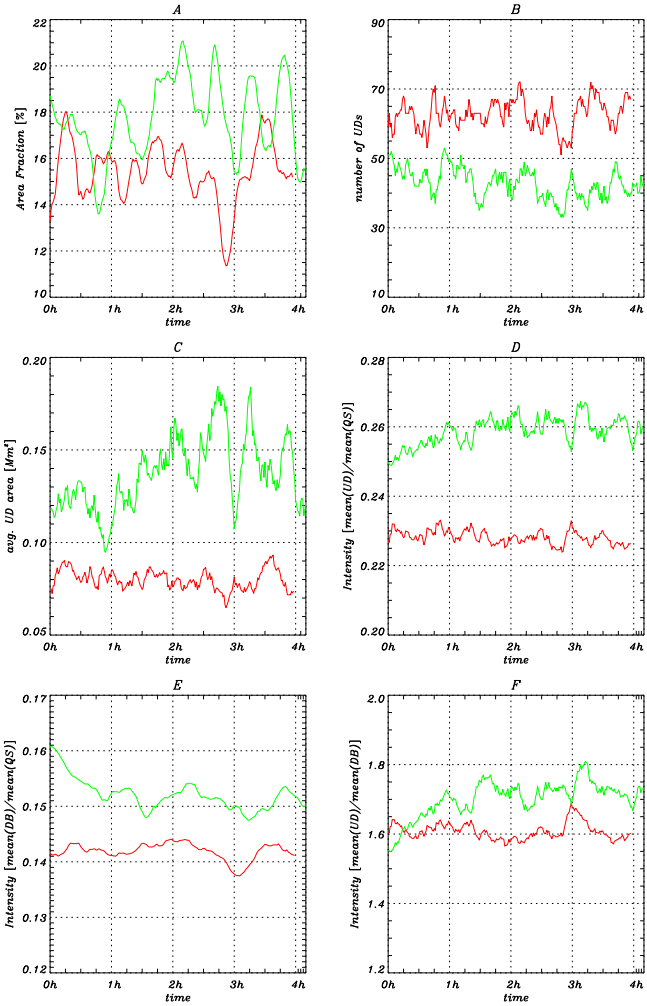}}
\caption{ Time evolution of UD properties, given separately
for the two phases P1 (cooling phase, red curves) and P2 (convective
phase, green curves). {\em A:} UD area fraction; {\em B:} number of UDs;
{\em C:} average UD area; {\em D:} mean bolometric UD intensity
(normalized by the mean quiet-Sun value); {\em E:} average bolometric
intensity of DB (normalized by the mean quiet-Sun
value); {\em F:} mean bolometric UD intensity (normalized by the
corresponding average for the DB). Time is counted from the start of the
respective dataset (0.5 h and 6.2 h after the start of the simulation
for P1 and P2, respectively.)}
\label{fig:UDtemporal}
\end{figure}

Fig.~\ref{fig:UDtemporal} shows, separately for P1 and P2, the temporal
evolution of various average quantities of the segmented UDs and of the
remaining domain area, the DB.  The area fraction of UDs (top left
panel) tends to be somewhat higher in P2 compared to P1; its relatively
strong fluctuations result from the relatively small size of the
computational box.  In P2, there are less (top right panel) but,
on average, larger (middle right panel) and brighter (middle
left and bottom left panel) UDs, as already indicated by the snapshots
shown in Fig.~\ref{fig:snapshots}.  The DB is also slightly brighter in
P2 (bottom left panel).  The mean bolometric intensity of the UDs
approaches roughly stationary values of about 0.23 (P1) and 0.26 (P2) of
the quiet-Sun value, or about 1.6 (P1) and 1.7 (P2), respectively, in
units of the corresponding DB intensity (bottom right panel).
These results indicate that the upflows underlying the UDs contribute
more strongly to the convective energy transport in P2. This includes not
only the fraction of the radiative output directly covered by
the UDs (about 30\%), but also the horizontal radiative losses of the
upflow plumes, which heat their environment and thus contribute to the
energy output of the DB. The higher average brightness of the DB in the
second phase (P2) is thus consistent with a bigger contribution of the
UDs to the overall convective energy transport.

The average UD area resulting from our simulation and segmentation
analysis is 0.08 Mm{$^2$} for P1 and 0.14 Mm{$^2$} for P2, corresponding
to average diameters of $320$~km and $420$~km under the assumption of
circular areas. These values are higher than those reported in recent
observational studies of UDs with the 1-m SST on La Palma:
\citet{Riethmueller:etal:2008b} give an average area of 0.04~Mm{$^2$},
\citet{Sobotka:Hanslmeier:2005} estimate 0.025~Mm{$^2$} while
\citet{Sobotka:Puschmann:2009} even find a value near 0.01~Mm{$^2$}.  On
the other hand, the UDs with dark lanes studied by the latter authors
have an average area around 0.08~Mm{$^2$} and even larger UDs have been
studied by \citet{Bharti:etal:2007b}. Also, \citet{Watanabe:etal:2009a}
find an average area of about 0.1~Mm{$^2$} from Hinode data. In such a
comparison one has to take into account that the lowest intensity
thresholds in our MLT segmentation are not far above the DB values,
while the observational analyses typically define the edges of an UD by
the half-width of the local intensity contrast or by the inflection
point of the intensity profile. These procedures tend to yield smaller
areas than our approach, which covers the full extension of the UDs
\citep[cf., for example, the UD outlines shown in Figs.~5 and 6
of][]{Riethmueller:etal:2008b}. We tested this conjecture by applying
the procedure of \citet{Riethmueller:etal:2008b} to a few representative
snapshots in our datasets. We found that the resulting UD areas are, on
average, about 50\% smaller than the values determined by our MLT
procedure. This brings the simulation and the observational results into
rough agreement -- the simulations possibly lacking a population of very
small, short-lived UDs \citep{Sobotka:Puschmann:2009}.

\begin{figure}
\centering
\resizebox{\hsize}{!}{\includegraphics{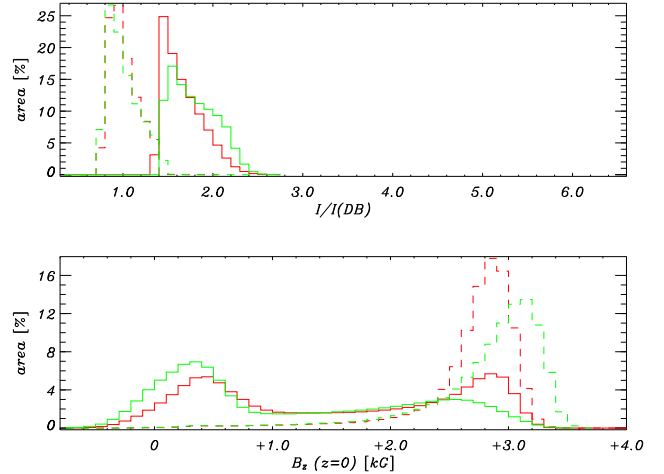}}
\caption{ Area histograms of bolometric intensity (upper
panel) and vertical field strength (lower panel). The field strength is
given taken at $z=0$, corresponding to the average geometrical height
level of Rosseland optical depth unity. Solid curves refer to UDs,
dashed curves to DB. Red and green curves indicate datasets P1 and P2,
respectively.}
\label{fig:hist_I_B_bol}
\end{figure}

\begin{figure}
\hspace{-3mm}
\centering
\resizebox{\hsize}{!}{\includegraphics{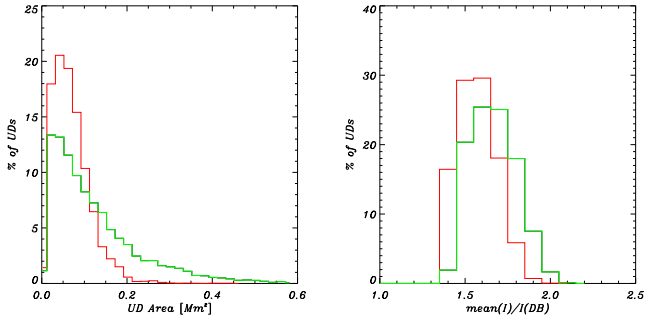}}
\caption{ Histograms of UD area (left) and mean bolometric
         brightness (right, normalized to the average DB)
         for all segmented maps.  Red and green curves refer to datasets
         P1 and P2, respectively.}
\label{fig:hist_allUD}
\end{figure}

For the full datasets P1 and P2, Fig.~\ref{fig:hist_I_B_bol} shows area
histograms of the bolometric intensity and the vertical component of the
magnetic field at constant height $z=0$, which corresponds to the
mean level of Rosseland optical depth unity. Histograms for
UDs and DB are given separately (including all image pixels belonging to
each class). The UD intensity distributions reflect the fact that UDs
tend to be brighter in P2.  The tail of DB intensities
overlapping with the UD distribution results from the `trenches'
generated by the MLT algorithm to separate closely neighboring or
clustered UDs and the exclusion of the faintest UDs.

The area histograms of the vertical magnetic field given in the lower
panel of Fig.~\ref{fig:hist_I_B_bol} shows a very broad distribution of
field strength in the UDs, ranging from slightly negative values up to
about 3000~G. While the field strength becomes very low owing to
expansion and flux expulsion in the core of the rising plumes that
generate the UDs, horizontal radiative heating extends the wings of the
bright intensity structure (i.e., the UD) into the surrounding umbra
with high field strength. Since we segment the images according to the
intensity structure, the peripheries of the so-defined UDs harbor
strong magnetic fields. Negative field strength arises when a strong
downflow catches and drags a magnetic field line, so that it is
stretched into a hairpin-like shape (see Fig.~\ref{fig:maps_B_v}
and Sec.~4.3). The average field strength in the DB is higher in P2
since the larger area fraction of UDs compresses the magnetic flux in
the DB.

The distributions of UD area and mean brightness for all segmented maps
(with one minute cadence, so that UDs are considered more than once, in
various stages of their evolution) are illustrated by the histograms
shown in Fig.~\ref{fig:hist_allUD}. UDs with areas below 0.02~Mm{$^2$}
were omitted as most of them represent fluctuations in the DB. The
area distribution for P2 is broader, with less smaller and more larger
UDs than for P1.  The largest UDs have areas of 0.44~Mm{$^2$} and
0.79~Mm{$^2$} for P1 and P2, respectively.  The distribution of average
UD bolometric brightness contrast (right panel of
Fig.~\ref{fig:hist_allUD}) is shifted towards higher values for P2, the
mean values being 1.62 (P1) and 1.72 (P2).

\begin{figure}
\centering
\resizebox{\hsize}{!}{\includegraphics{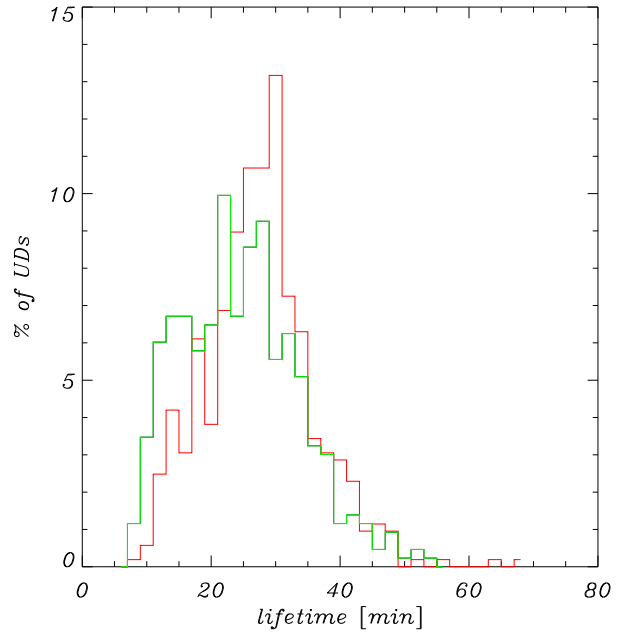}}
\caption{ Histogram of lifetimes for UDs that could be
         tracked during their whole evolution.  Red and green curves
         refer to datasets P1 and P2, respectively.}
\label{fig:hist_lifetime}
\end{figure}

Since the MLT tracking algorithm identifies and tracks UDs during their
whole life cycle, we determined lifetimes for 558 UDs from P1 and
500 UDs from P2. The lifetime distributions are shown in the form of
histograms in Fig.~\ref{fig:hist_lifetime} (excluding the roughly 10\%
of the UDs with lifetimes below 8 minutes, which seem to constitute a
separate group of low-amplitude fluctuations). The lifetime
distributions are similar for both datasets, with average values of
28.2~min (P1) and 25.1~min (P2).  The mean lifetimes reported from
observations range from 3~min to about an hour, with recent
high-resolution observations typically indicating rather short mean
lifetimes below 10 minutes \citep[e.g.,][]{Kusoffsky:Lundstedt:1986,
Ewell:1992, Sobotka:etal:1997, Kitai:etal:2007, Riethmueller:etal:2008b,
Sobotka:Puschmann:2009, Watanabe:etal:2009a}. However, similar to our
simulation results, the distribution of lifetimes from any observation
is rather broad and `typical' lifetimes cannot be defined easily
\citep{Socas-Navarro:etal:2004}. Taken at face value, the observations
indicate the existence of a large number of short-lived UDs that are not
shown by the simulations. However, it is also possible that some
observational lifetimes are affected by brightness fluctuations of UDs
and by seeing effects.

\begin{figure}
\centering
\resizebox{\hsize}{!}{\includegraphics{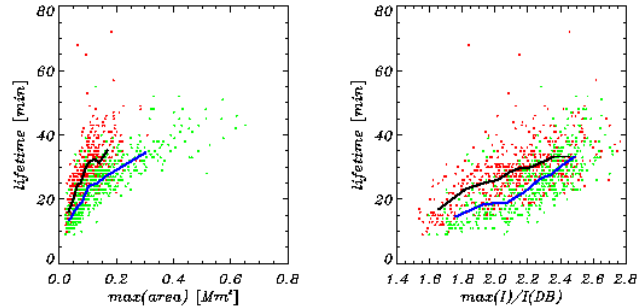}}
\caption{ Relationship between UD lifetime, area and
         brightness.  {\em Left:} scatter plot of lifetime
         vs. maximum area; {\em Right:} scatter plot of lifetime
         vs. maximum area-averaged bolometric brightness reached during
         the lifetime of an UD (normalized by the average brightness of
         the DB).  Red and green dots refers to datasets P1 and
         P2, respectively; solid black (P1) and blue (P2) lines connect
         binned values for 50 points each.}
\label{fig:scatter_lifetime}
\end{figure}

\begin{figure}
\centering
\resizebox{\hsize}{!}{\includegraphics{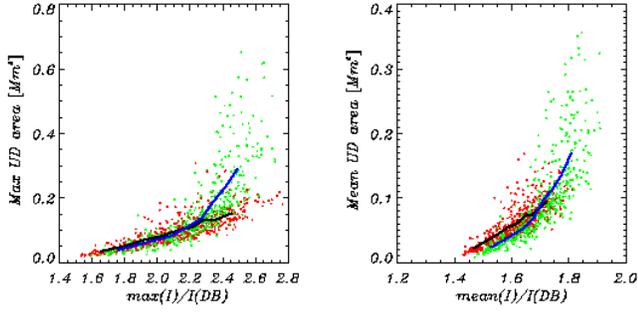}}
\caption{ Relation between area and brightness for the same
  set of UDs as in Fig.~\ref{fig:scatter_lifetime}. {\em Left:}
  scatter plot of maximum area vs. maximum (normalized) bolometric
  brightness during the lifetime of an UD; {\em Right:} scatter plot of
  mean UD area (over UD lifetime) vs. mean area-averaged brightness
  (over UD lifetime).  Red and green color refers to datasets P1 and P2,
  respectively; solid black (P1) and blue (P2) lines connect binned
  values for 50 points each.}
\label{fig:scatter_lifetime2}
\end{figure}

The relationship between lifetime and maximum UD area and brightness in
the course of the UD evolution is shown in
Fig.~\ref{fig:scatter_lifetime} in the form of scatter plots with curves
connecting binned values. Although there is a significant amount of
scatter, the plots indicate that longer-lasting UDs tend to be larger
and brighter. A qualitatively similar lifetime-size relation is reported
by \citet{Riethmueller:etal:2008b} for UDs with lifetimes below 20 min
while longer-lived UDs are not found to be systematically larger.

The lifetime-area and lifetime-brightness correlations are probably due
to the fact that stronger and more extended convective upflows are
maintained longer and create larger and brighter UDs: higher upflow
speed entails a bigger convective energy flux density (brighter UDs) and
also more mass flux and kinetic energy available for the sideways
expansion of the upflow plume (bigger UDs).  This explanation is in line
with the relationship between UD area and (maximum and mean) brightness
shown in Fig.~\ref{fig:scatter_lifetime2}. Similar trends were
found observationally by \citet{Tritschler:Schmidt:2002} and by
\citet{Sobotka:Puschmann:2009} while \citet{Riethmueller:etal:2008b} do
not find a clear relationship.

\subsection{UD properties and optical-depth dependence at 630 nm}
\label{subsec:prop630}

\begin{figure}
\centering
\resizebox{\hsize}{!}{\includegraphics{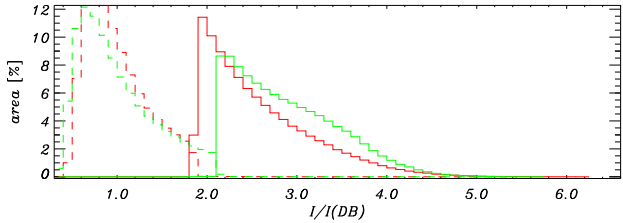}}
\caption{ Area histograms of the continuum intensity at 630
         nm. Solid curves refer to UDs, dashed curves to DB. Red and
         green curves indicate datasets P1 and P2, respectively.}
\label{fig:hist_630}
\end{figure}

\begin{figure}
\centering
\resizebox{\hsize}{!}{\includegraphics{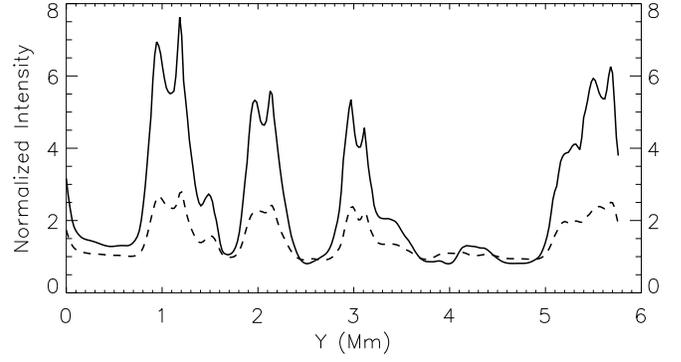}}
\caption{ Profiles of vertically emerging intensity along a
         cut at $x=4.6$~Mm through the image shown in the right panel of
         Fig.~\ref{fig:snapshots}. The solid line refers to the
         continuum intensity at 630~nm, the dashed line to the
         bolometric intensity.  Intensities are normalized individually
         by the corresponding local DB intensities.}
\label{fig:Icomp}
\end{figure}

The iron line pair near 630 nm and the nearby continuum is often used
for observations of sunspot fine structure.  In order to compare with
observational results derived from such data, we considered 57 (P1) and
74 (P2) snapshots from our simulation and calculated images of the
(vertically emerging) continuum intensity at 630~nm. These images were
segmented by the MLT procedure to obtain maps of UDs. In addition to the
UD properties studied in the previous section, we correlated the UD
maps with the distributions of vertical magnetic field and vertical
velocity on surfaces of equal optical depth at 630 nm.  A full synthesis
of line profiles and Stokes parameters for a detailed comparison with
observed spectro-polarimetric data is beyond the scope of this work and
will be carried out in a forthcoming paper (Vitas, V{\"o}gler \& Keller,
in preparation).

\begin{figure*}
\centering
\resizebox{\hsize}{!}{\includegraphics{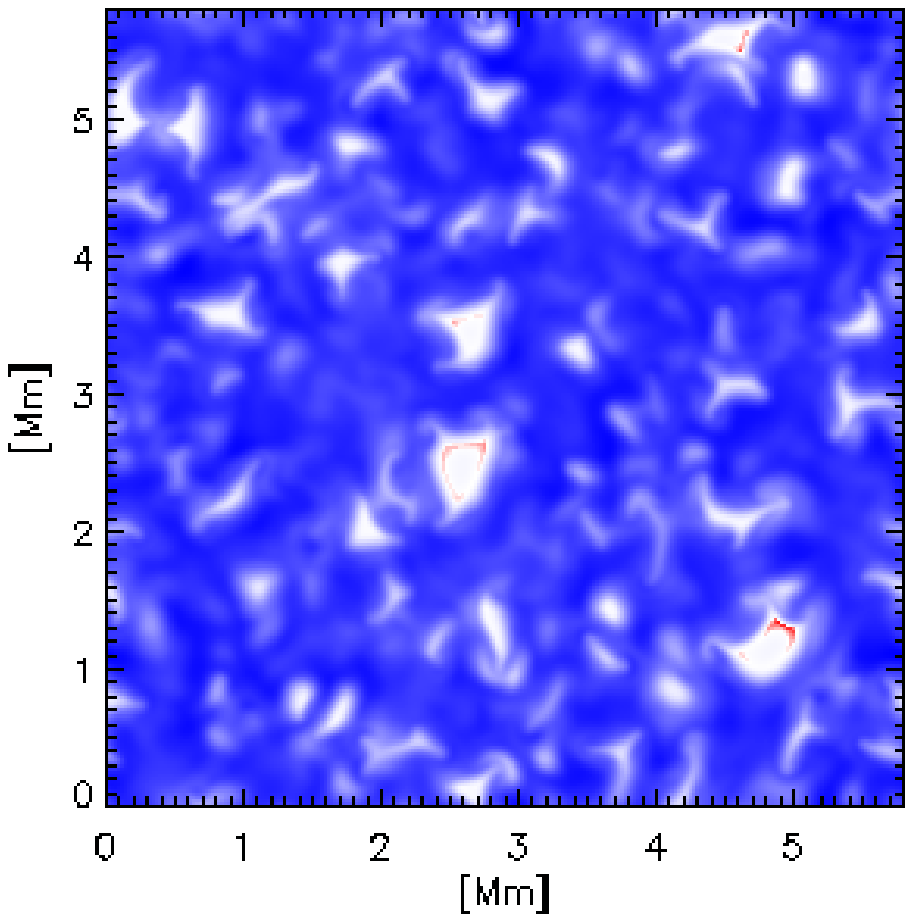}
                      \includegraphics{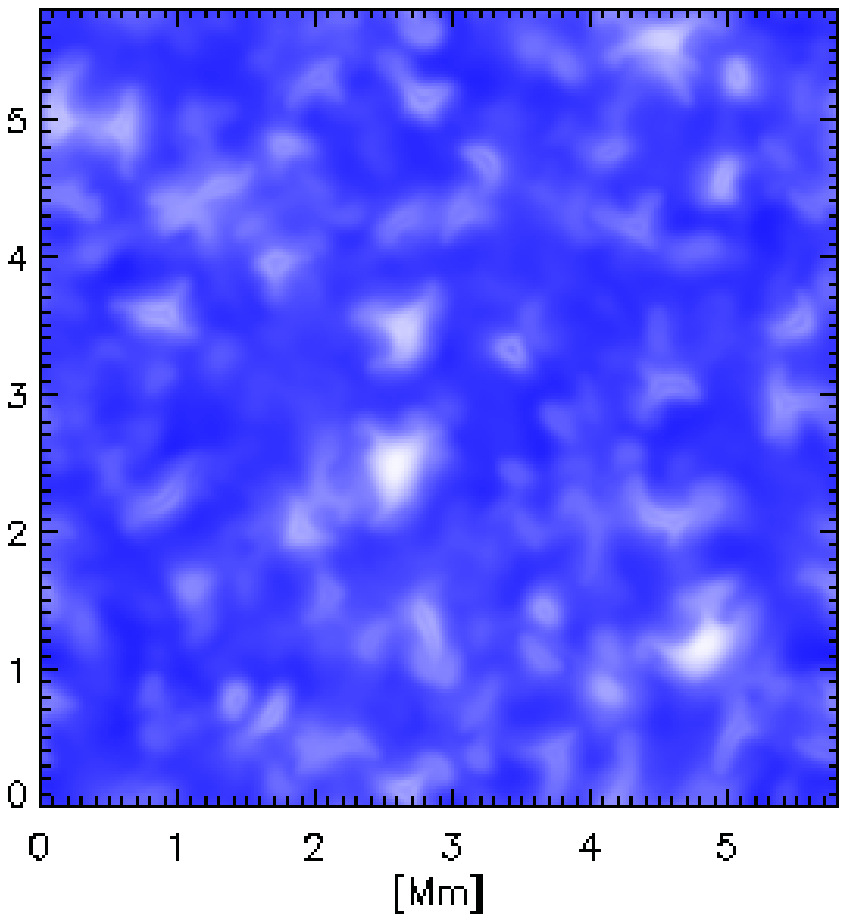}
                      \includegraphics{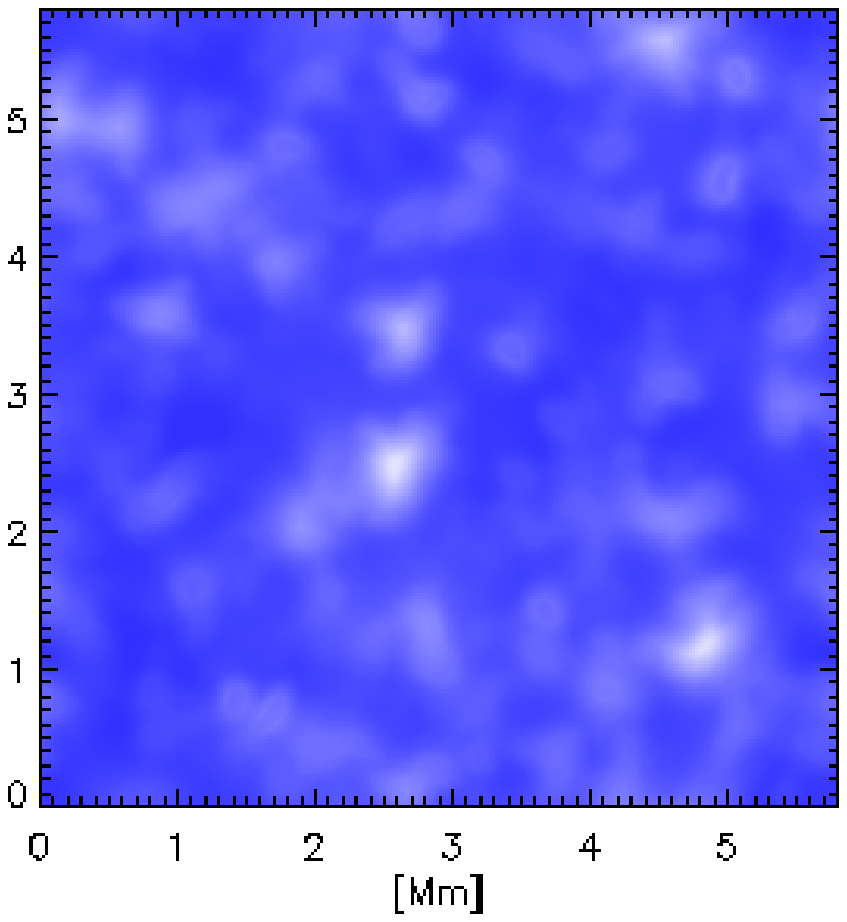}}
\resizebox{\hsize}{!}{\includegraphics{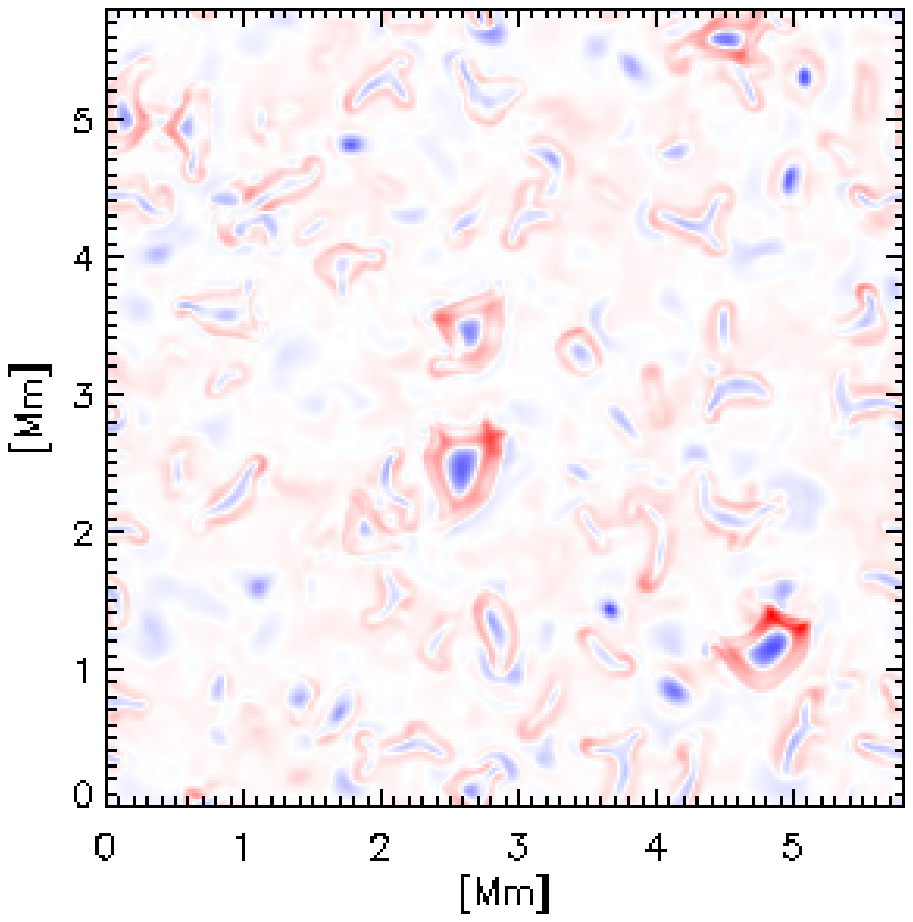}
                      \includegraphics{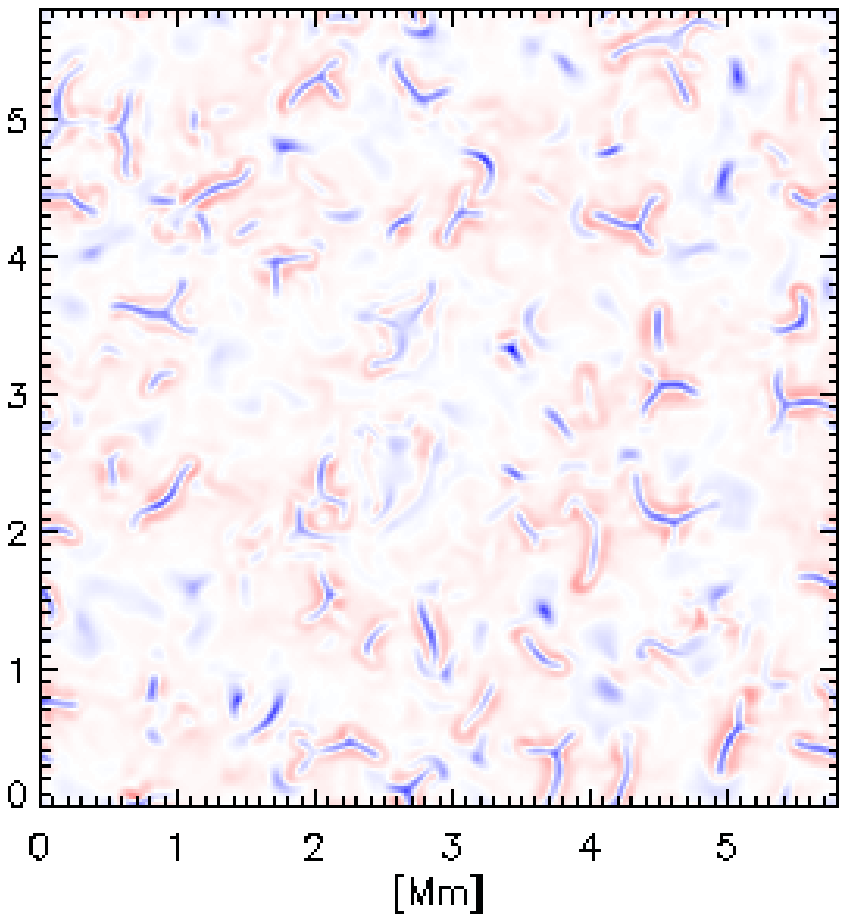}
                      \includegraphics{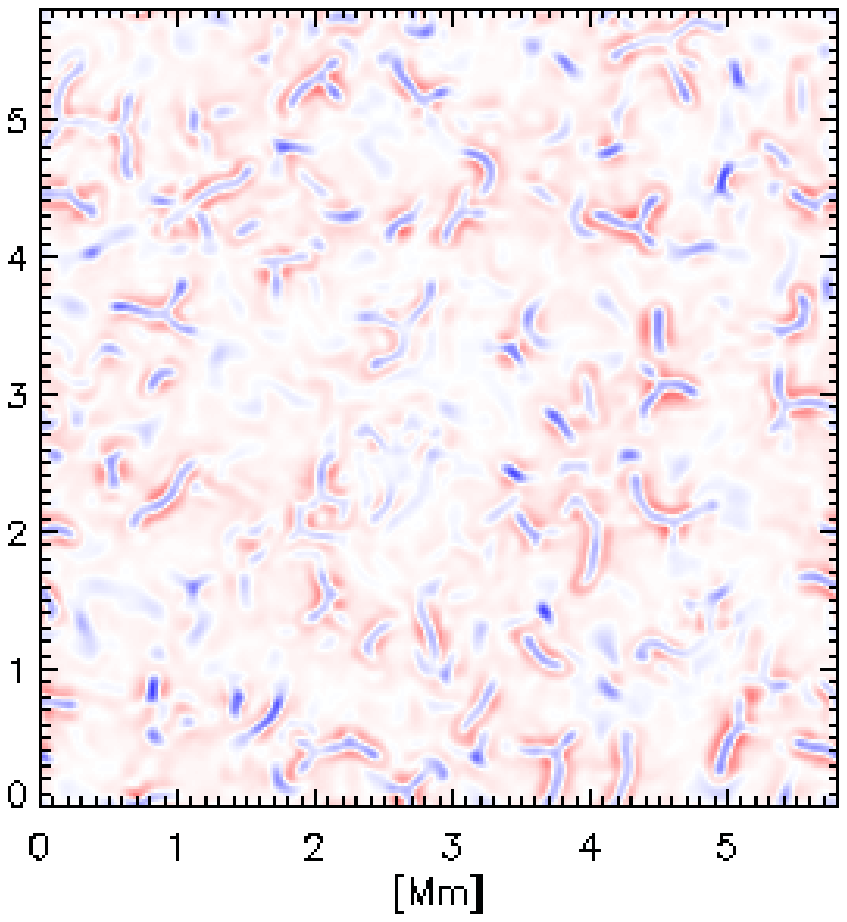}}
\caption{Maps of vertical magnetic field strength (upper row) and
  vertical velocity (lower row) on surfaces of constant optical depth at
  630~nm (from left to right: $\tau_{630}=1, 0.1, 0.01$, respectively).
  The color table for the magnetic field ranges from 3780~G (blue)
  over zero (white) to $-190$~G (red). For the velocity the range is
  from 1.9~km$\,$s$^{-1}$ (blue, upflow) over zero (white) to
  $-1.7$~km$\,$s$^{-1}$ (red, downflow).}
\label{fig:maps_B_v}
\end{figure*}

Area histograms for the continuum intensity at 630~nm are shown in
Fig.~\ref{fig:hist_630}. Compared to the histograms of the bolometric
intensity shown in Fig.~\ref{fig:hist_I_B_bol}, the intensity contrasts
between UDs and DB is much higher, resulting from the fact that 630 nm
is on the blueward side from the maximum of the respective Planck
functions. The significantly higher contrast between DB and UD is also
illustrated by the intensity profiles shown in Fig.~\ref{fig:Icomp}.
The mean UD intensity values in Fig.~\ref{fig:hist_630} are 2.58
for P1 and 2.88 for P2 (both relative to the corresponding average
DB). These values are consistent with the intensity ratios reported from
observations at 602~nm \citep[e.g.,][]{Sobotka:Hanslmeier:2005,
Sobotka:Puschmann:2009}.  Individual UDs in the simulation can
reach significantly higher intensities (see Fig.~\ref{fig:Icomp}).

\begin{figure}
\centering
\resizebox{\hsize}{!}{\includegraphics{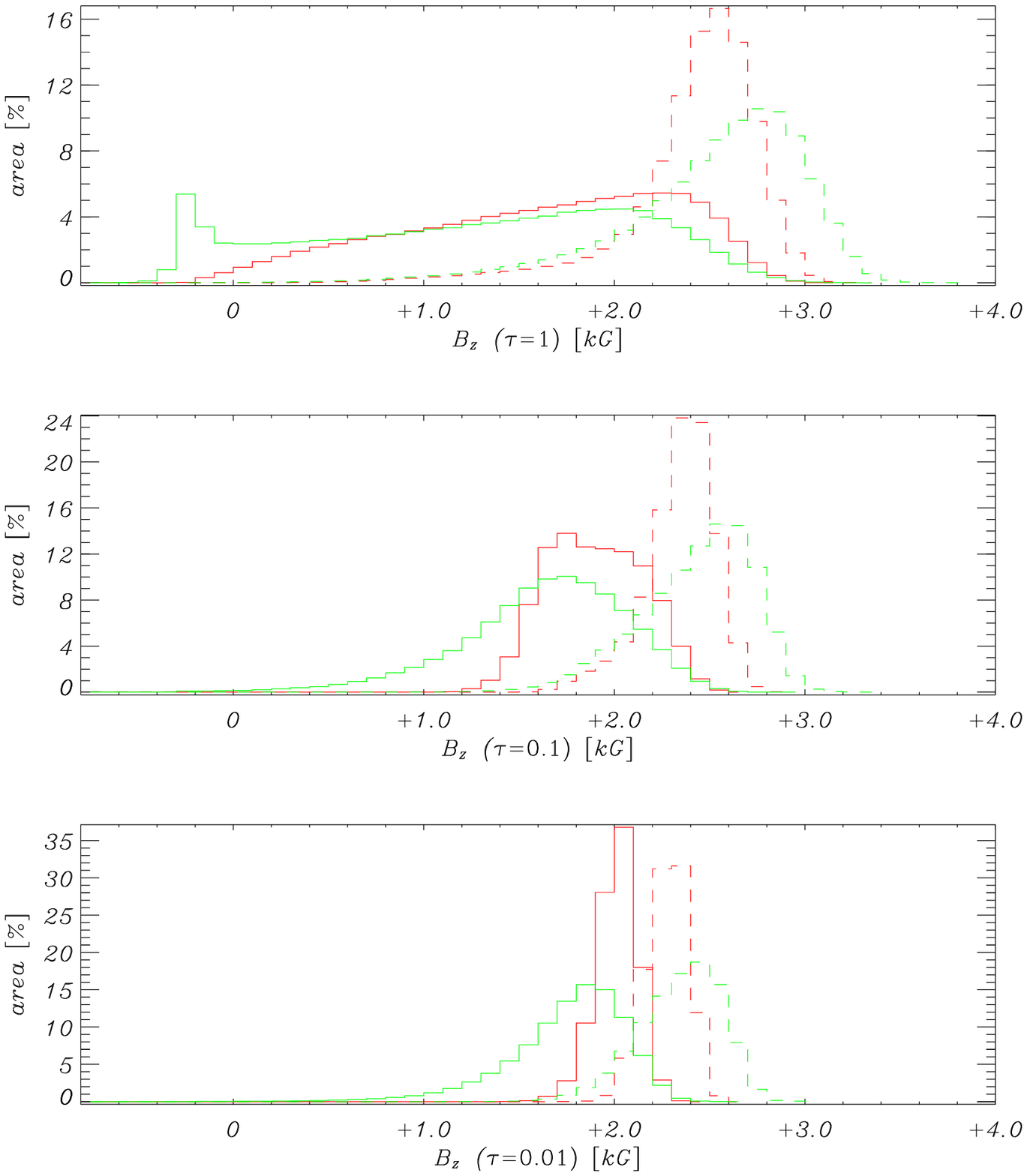}}
\caption{ Area histograms of the vertical magnetic field
         component on surfaces of constant optical depth at 630~nm.
         {\em Upper panel:} $\tau_{630}=1$; {\em middle panel:}
         $\tau_{630}=0.1$; {\em lower panel:} $\tau_{630}=0.01$.  Solid
         lines refer to UDs, dashed lines to the DB. The red and green
         colors indicate datasets P1 and P2, respectively. }
\label{fig:hist_630_mag}
\end{figure}

\begin{figure}
\centering
\resizebox{\hsize}{!}{\includegraphics{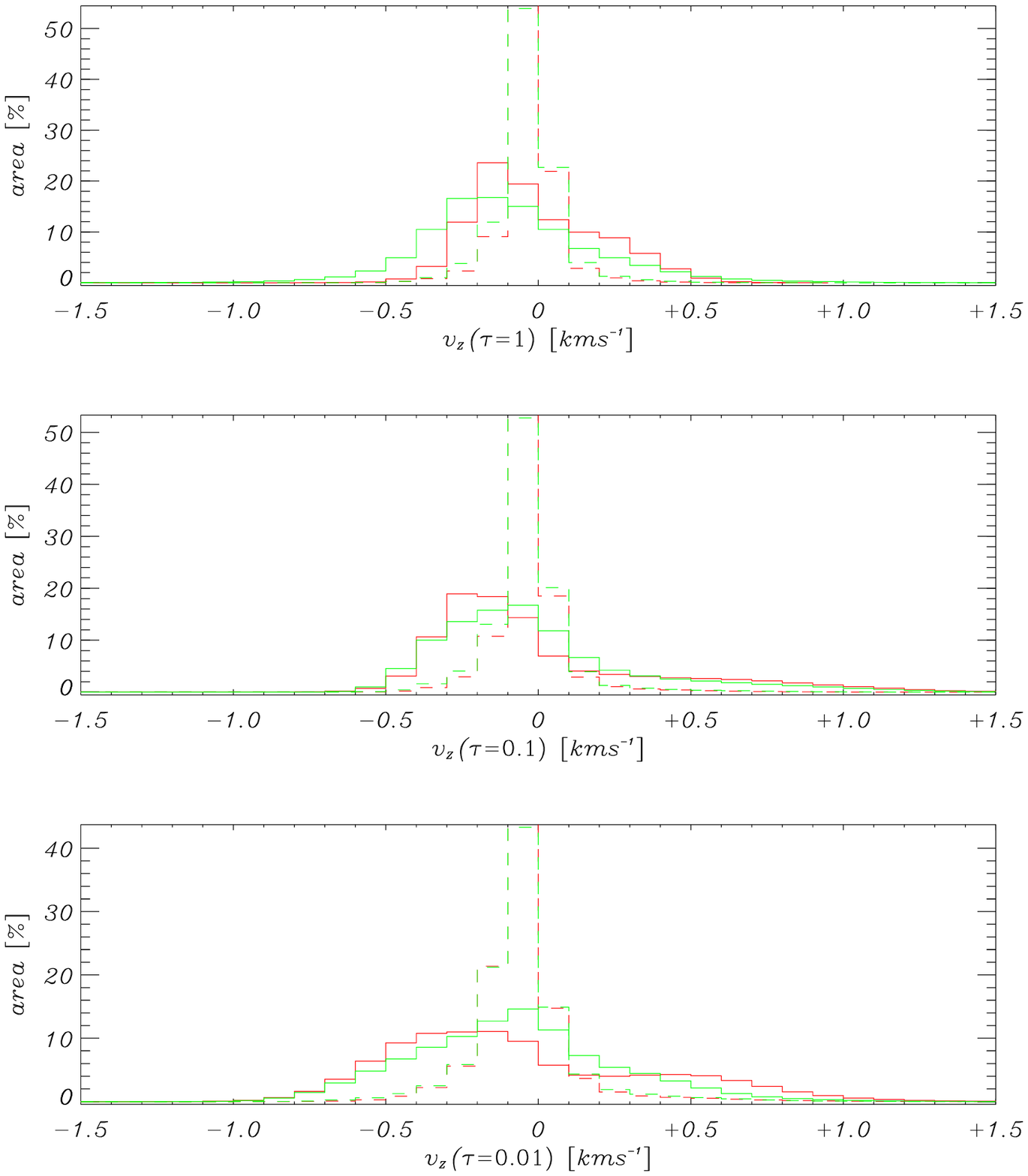}}
\caption{ Same as Fig.~\ref{fig:hist_630_mag} but for the
         vertical component of the velocity. Positive velocity
         corresponds to upward motion, negative velocity to a downward
         flow.}
\label{fig:hist_630_vel}
\end{figure}

Maps of the vertical magnetic field and vertical velocity on surfaces of
equal optical depth at 630~nm are given in
Fig.~\ref{fig:maps_B_v}. While the magnetic field distribution becomes
more diffuse and homogeneous with height, the velocity remains rather
intermittent. Strong downflows at the edges of the largest UDs at
$\tau_{630}=1$ (left panels) can capture inclined magnetic field lines
and drag them downwards, thereby creating a hairpin-like structure with
a patch of reversed polarity, indicated by red color in the
magnetic-field map. Quantitative information on the distributions is
provided by the area histograms for the same optical-depth levels shown
in Fig.~\ref{fig:hist_630_mag}. The histograms at $\tau_{630}=0.01$
roughly represent the results that would be obtained by carrying out
inversions of spectro-polarimetric observations in the neutral iron lines
at 630.15~nm and 630.25~nm. The histograms at $\tau_{630}=1$ are roughly
similar to those at constant geometrical depth $z=0$ shown in the lower
panel of Fig.~\ref{fig:hist_I_B_bol}, with a broad range of values in
the UDs.  At $\tau_{630}=0.1$, field strengths below 1000~G are rarely
found in UDs (particularly for those from P1) and at $\tau_{630}=0.01$,
the UD distributions are further shifted to higher field
strengths. For instance, 90\% of the UDs of dataset P1 have field
strengths between 1800~G and 2200~G at $\tau_{630}=0.01$. This effect is
due to the elevation of the surfaces of constant optical depth in UDs
(in comparison to the DB) and the cusp-like shape of the UD structure
\citep[cf.][]{Schuessler:Voegler:2006, Riethmueller:etal:2008a,
Bharti:etal:2009}. The larger, stronger UDs in the convective phase
(dataset P2, green lines) reach higher into the umbral atmosphere
[cf. Fig.~\ref{fig:scat_B_v_z}] and show therefore still a stronger
signature of the reduced field strength at $\tau_{630}=0.01$ than the
UDs from P1. For both datasets, the distributions for UDs and DB
approach each other as the optical depth decreases; this reflects the
decreasing area fraction of the cusp-shaped UD structures with
increasing height.

The corresponding area histograms of the vertical velocity given in
Fig.~\ref{fig:hist_630_vel} show rather weak motion in the DB: at all
three $\tau$ levels, about 50 $\%$ of DB shows almost no vertical flow
while the rest of the area has velocities in the range $\pm$
$200\,$m$\,$s$^{-1}$. The strong upflows creating the UDs have already
been considerably braked when they reach the level
$\tau_{630}=1$. Higher up at levels $\tau_{630}=0.1$ and
$\tau_{630}=0.01$, high-velocity tails (up to $\sim \pm 1\,$km$\,$s$^{-1}$)
appear in the distributions. These are mostly due to the upward-directed,
jet-like `valve flows' from the UD cusps \citep{Choudhuri:1986,
Schuessler:Voegler:2006} and the corresponding return flows. The latter,
however, may be affected by the presence of the closed upper boundary
of the simulation box.

\begin{figure}
\centering
\resizebox{1.0\hsize}{!}{\includegraphics{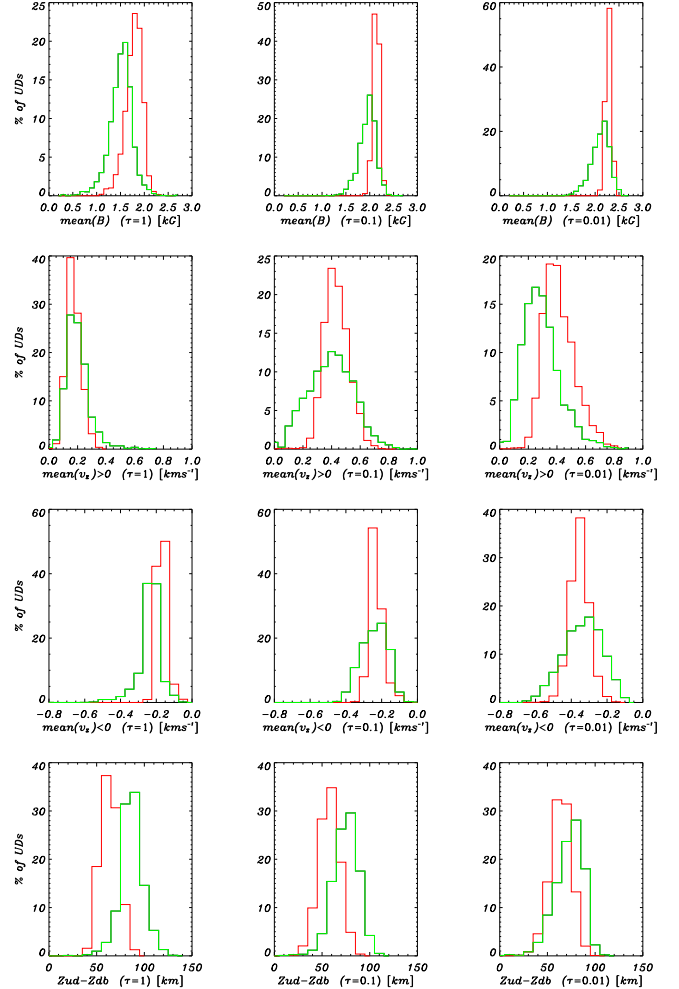}}
\caption{ Distributions of averages over UD area of vertical
    magnetic field (top row) and vertical velocity (separately for
    upflows, $v_z>0$, and downflows, $v_z<0$ in the second and third
    row, respectively), all on surfaces of constant optical depth at
    630~nm (left column: $\tau_{630}=1$, middle column:
    $\tau_{630}=0.1$, right column: $\tau_{630}=0.01$).  The bottom row
    gives the elevation of the optical-depth levels of the UDs with
    respect to the corresponding mean levels in the DB.  Red and green
    lines refer to UDs from datasets P1 and P2, respectively.}
\label{fig:histograms_B_v_z}
\end{figure}

\begin{figure}
\centering
\resizebox{1.0\hsize}{!}{\includegraphics{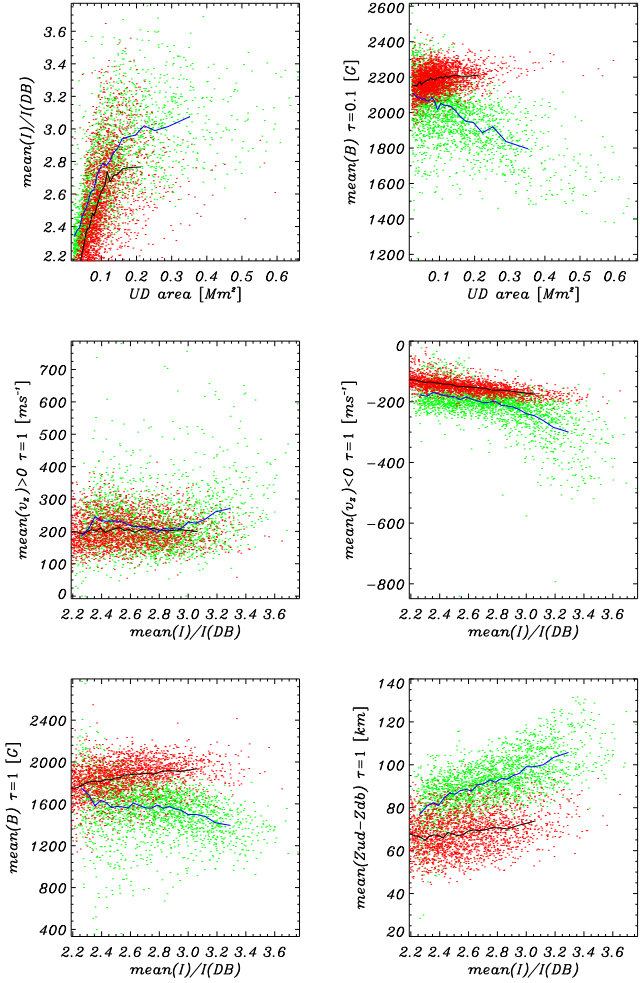}}
\caption{ Relationship between properties of UDs defined by
  segmentation of continuum images at 630~nm. Mean quantities are
  averages over the area of the individual UDs. Red and green dots
  refer to datasets P1 and P2, respectively; solid black (P1) and blue
  (P2) lines connect binned values for 100 points each. Positive velocity
         corresponds to upward motion, negative velocity to a downward
         flow.}
\label{fig:scat_B_v_z}
\end{figure}

Figure~\ref{fig:histograms_B_v_z} illustrates, by means of histograms,
the properties of average vertical magnetic field and velocity at
various optical depths, together with the average elevations of the
optical-depth levels for all UDs from the datasets P1 and P2. The
magnetic field histograms show that, at all optical-depth levels, the
expulsion of magnetic flux from the expanding upflow plumes underlying
the UDs is more clearly reflected in the case of P2.  However, in most
cases the region of very low field strengths does not extend above the
surface of $\tau_{630}=1$.  The cusp shape of the vertical UD structure
\citep[see Fig.~2 of][]{Schuessler:Voegler:2006} leads to an increase of
the field strength for the iso-$\tau$ surfaces above $\tau_{630}=1$.  In
addition, the UD area is defined by the intensity structure
corresponding to a height near $\tau_{630}=1$; the cusp's cross section
shrinks toward higher iso-$\tau$ surfaces, so that we progressively
sample more of the strong-field region surrounding the plume.  A similar
situation is found for the upflows (second row), where the average
velocities do not exceed a few hundred m$\,$s$^{-1}$ at
$\tau_{630}=1$. At higher levels, the jet-like outflows along the cusps
of the upflow plumes are seen. The distribution of upflows is shifted
towards lower velocities for the UDs from P2, since bigger and flatter
UDs show weaker outflows. This results from the fact that the height
reached by the upflow plume is only weakly dependent on its size (since
the stratification above optical depth unity is strongly subadiabatic),
so that the cusp overlying a large UD is broader (has a larger aspect
angle) and the whole structures thus is flatter.  Therefore, the channelling
of the upflowing matter into the top of the cusp is less efficient and
the jet-like outflows are weaker.
The downflows (third row) at the UD
periphery (for $\tau_{630}=1$) and adjacent to the upflow jets (for the
other levels) are similar in average magnitude for P1 and P2 while the
distributions for P2 are somewhat broader.

The bottom row of Fig.~\ref{fig:histograms_B_v_z} shows histograms of
the relative height difference between the levels of constant optical
depth in UDs (averaged over UD area) and in the mean DB, thus
representing the elevation of the optical-depth levels in the UDs
relative to the DB. While the mean elevations range between 60--90~km,
the levels for the UDs from P2 are, on average, 10--20~km higher than
the corresponding levels of UDs from P1. Thus, the P2 UDs tend to penetrate
somewhat higher into the umbral photosphere than their counterparts from
P1.

The elevation of the surfaces of constant optical depth above the upflow
plumes tends to hide their magnetic and flow structure from
spectroscopic observations \citep[e.g.,][]{Degenhardt:Lites:1993a,
Degenhardt:Lites:1993b}. It is therefore not surprising that
observational results for the magnetic and flow properties of UDs do not
provide a unique picture. A reduced field strength in UDs has been
repeatedly reported \citep[e.g.,][]{Wiehr:Degenhardt:1993,
Socas-Navarro:etal:2004} and some authors also find indications for a
decrease of field strength with depth \citep{Riethmueller:etal:2008a,
Bharti:etal:2009}. Other studies \citep{Sobotka:Jurcak:2009,
Watanabe:etal:2009b} do not show significantly lower field strengths
in UDs.  While upwellings in the deep layers of UDs were found
\citep{Socas-Navarro:etal:2004, Rimmele:2004, Riethmueller:etal:2008a,
Bharti:etal:2007a, Bharti:etal:2009}, the jet-like upflows predicted by
the simulations in the higher layers have not been observed so far. This
may be due to the fact that spectroscopic studies mostly refer to large
UDs, for which the jets are weaker or absent.

The relations between various mean quantities for the full set of UDs
found by segmentation of continuum images at 630~nm are illustrated by
scatter plots in Fig.~\ref{fig:scat_B_v_z}. The positive correlation
between UD area and brightness seen in Fig.~\ref{fig:scatter_lifetime2}
(for bolometric brightness) is confirmed also for 630~nm (top left
panel). Although there is a large amount of scatter, UDs from P2 tend to
be brighter for the same area, possibly reflecting higher upflow speeds
(below optical depth unity) with a higher convective energy flux.

This interpretation is supported by the decrease of the magnetic
field strength with increasing UD area at $\tau_{630}=0.1$ (top right
panel) and with increasing intensity at $\tau_{630}=1$ (bottom left
panel) for P2 as well as by the rise of the elevation of the level
$\tau_{630}=1$ with increasing UD intensity (bottom right panel).  The
downflows at $\tau_{630}=1$ intensify with increasing brightness (middle
right panel). The upflows $\tau_{630}=1$ are largely independent of
brightness (middle left panel); since the level surface is higher for
brighter UDs, this also indicates stronger underlying upflows.

\section {Conclusions}

The simulation of magneto-convective energy transport in a strong
vertical magnetic field exhibits narrow upflow plumes, whose properties
can be compared with observations of (central) umbral dots in
sunspots. Using a multi-level image segmentation and tracking algorithm,
we analyzed sets of simulated UDs from two phases of the
simulation, the thermal relaxation phase (P1) and the quasi-stationary
phase (P2). This led to the following results:
\begin{enumerate}
\item {\em Size:} Histograms of UD area indicate that there is no
      `typical' size of the simulated UDs. The average area is
      0.08~Mm$^2$ in P1 and 0.14~Mm$^2$ in P2. Reported values from
      recent high-resolution observations are somewhat lower. This
      is partly due to different size definitions, but could also
      indicate a lack of small, short-lived UDs in the simulation.
\item {\em Brightness:} Averaged over their area, the bolometric
      brightness of the simulated UDs exceeds the surrounding dark
      background by a factor 1.6 (P1) and 1.7 (P2), respectively. The
      corresponding values for the continuum brightness at 630~nm
      are 2.6 (P1) and 2.9 (P2). For the peak intensities, the factors
      can reach markedly higher values between about 3 (bolometric) and
      8 (630 nm). None of the simulated UDs exceeds the corresponding
      brightness values of the quiet Sun.  Comparison with observations
      is complicated by the often unknown amount of straylight
      contamination.
\item {\em Lifetime:} Average lifetimes for the simulated UDs are 28~min
      (P1) and 25~min (P2). Reported lifetimes from observations vary
      significantly, but recent high-resolution data typically indicate
      shorter lifetimes, although long-lived UDs are also found. It is
      possible that the simulations, owing to limitations in spatial
      resolution, miss a population of small, short-lived UDs.
\item {\em Correlations:} Larger UDs tend to be brighter and live
      longer, although there is a significant amount of scatter.
      Similar trends have been reported from observations.
\item{\em Magnetic field and flows:} The drastic reduction of the
      magnetic field and the strong flows in the near-photospheric parts
      of the upflow plumes are largely hidden from spectro-polarimetric
      observations. This is caused by the elevation of the surfaces of
      constant optical depth unity, which bulge upward over columns of
      hot rising plasma. Consistent with observational results, only a
      moderate field reduction and weak flow signatures are
      expected at optical-depth levels where relevant photospheric lines
      are formed.
\end{enumerate}
In summary, the comparison of the properties of simulated and observed
(central) UDs indicates that the simulations have caught the basic
underlying mechanism for the formation of these bright
structures. Differences in detail are not surprising, given the still
unsufficient spatial resolution of both simulations and observations.
Simulations of full sunspots, which have recently become available
\citep{Rempel:etal:2009b}, show that the similar magneto-convective
processes are responsible for the formation of umbral dots and penumbral
filaments, also clarifying the relationship between central and
peripheral UDs. More comprehensive parameter studies are needed to
reveal the dependence of UD properties on background field strength,
spatial resolution, and vertical extension of the computational
box. Also, the calculation of synthetic Stokes profiles will permit a
direct comparison with spectro-polarimetric observations and inversions.


\bibliographystyle{aa}
\bibliography{13328.bbl}

\end{document}